\documentclass[journal,10pt,comsoc]{IEEEtran}
\usepackage{blindtext, graphicx}
\usepackage{amsmath}

\makeatletter
\def\ps@IEEEtitlepagestyle{
  \def\@oddfoot{\mycopyrightnotice}
  \def\@evenfoot{}
}
\def\mycopyrightnotice{
  {\footnotesize
  \begin{minipage}{\textwidth}
  \centering
Copyright (c) 2015 IEEE. Personal use of this material is permitted. However, permission to use this material for any other purposes
must be obtained\\
from the IEEE by sending a request to pubs-permissions@ieee.org.
  \end{minipage}
  }
}
\usepackage{tikz}

\DeclareMathOperator*{\argmin}{arg\,min}
\usetikzlibrary{shapes,arrows}
\usepackage{array, tabularx}
\usepackage{multirow,multicol, makecell, booktabs, caption}

\usepackage{pifont}
\usepackage{array,tabularx}
\usepackage{lipsum}
\usepackage{xcolor}
\usepackage{caption}
\usepackage{subcaption}
\usepackage{bbm}
\newcolumntype{P}[1]{>{\centering\arraybackslash}p{#1}}
\newcolumntype{C}[1]{>{\centering\arraybackslash}m{#1}}
\usepackage{amstext}  
\usepackage{footnote}
\makesavenoteenv{tabular}
\makesavenoteenv{table}

\usepackage{times}
\usepackage{fancyhdr,graphicx,amssymb}
\usepackage[linesnumbered,ruled,vlined]{algorithm2e}

\usepackage{float}
\floatstyle{plaintop}
\restylefloat{table}

\SetCommentSty{mycommfont}
\include{pythonlisting}
\begin{document}
\title{ Large Kernel Polar Codes with efficient Window Decoding}

\author{\IEEEauthorblockN{Fariba Abbasi, \textit{Student Member, IEEE} and Emanuele Viterbo, \textit{Fellow, IEEE}
\thanks{The authors are with the Department of Electrical and Computer Systems Engineering (ECSE), Monash University, Melbourne, VIC3800,
Australia. E-mail:
\{fariba.abbasi, emanuele.viterbo\}@monash.edu.
These authors’ work was supported by the Australian Research Council under
Discovery Project ARC DP160100528.
}\\}
}


\maketitle

\begin{abstract}
\textcolor{black}{Window decoding is useful for decoding polar codes defined by kernels that do not have Ar\i kan's original form. We modify 
arbitrary polarization kernels of size $2^t \times 2^t$ to reduce the time complexity of window decoding.}
This modification is based on the permutation of the columns of the kernel.
This method is applied to some of the kernels constructed in the literature of size $16$ and $32$, with different error exponents and scaling exponents such as eNBCH kernel. It is shown that this method reduces the complexity of the window decoding significantly without affecting the performance.

\end{abstract}



\section{Introduction}
Polar codes, introduced by Ar\i kan \cite{Arikan}, are the first family of capacity-achieving codes with low complexity successive cancellation (SC) decoder. 
However, the performance of the polar codes at finite block lengths is not comparable with the state of art codes, due to (i) sub-optimality of the SC decoder and (ii) imperfectly polarized bit-channels resulted from $2 \times 2$ Ar\i kan's polarization kernel.
To solve the second problem, \cite{Korada} first  proposed to replace Arikan's kernel 
with a larger matrix with a better polarization rate. 
Later, different binary and non-binary linear and non-linear kernels with different polarization properties have been constructed in \cite{Mori}-\cite{Lin}. 
However, applying the classical SC decoder to these kernels is not practical, due to high decoding complexity. 
In \cite{TrifonovRS}, Trifonov used a window decoder for polar codes constructed with non-binary Reed-Solomon (RS) kernels \cite{Mori}.
\textcolor{black}{In \cite{T16} the window decoder, offering  further complexity reduction, was applied on some large binary kernels.} The Window decoder exploits the relationship between arbitrary kernels and the Ar\i kan's kernel. \textcolor{black}{However, the complexity of the window decoder for any arbitrarily large kernel (e.g. eNBCH kernels, \textcolor{black}{\cite{Vera}, \cite{TNew}}) is too high for practical implementation.}

A heuristic construction was proposed in  \cite{T32} for binary kernels of dimension $16$, which minimizes the complexity of the window decoder  and achieves the required rate of polarization. The authors achieved these goals by applying some elementary row operations on Ar\i kan's kernel. \textcolor{black}{However, a systematic design of large kernels with the required polarization properties, which admit low-complexity decoder, is still an open problem.}

\textcolor{black}{In this paper, we modify some arbitrary $2^t \times 2^t$ large kernels with good prolarization rates, to reduce the complexity of the window decoder.} This modification is based on the search through $(2^t)!$ column permutations of the kernel which do not affect its polarization properties \cite{Fazeli}. Since exploring all possible permutations is not practical, we propose a sub-optimal algorithm to reduce the search space significantly and find good column permutations independently of the structure of the original kernel. Then, we apply our algorithm to the kernels of size $16$ and $32$ constructed in the literature with higher error exponents. 

\textcolor{black}{Our proposed algorithm is independent of the structure of the original kernel and can be used to tackle the first and second open problems. The important contribution of our algorithm is that it can systematically be applied to any arbitrary kernel with a good polarization rate, and it significantly reduces the complexity of the window decoder.} 

Recently, in an independant work, authors in \cite{TNew} also suggested a suboptimal search algorithm of column permutations to reduce the complexity of the Viterbi algorithm for eNBCH kernels. This algorithm specifically takes advantage of eNBCH structure for reducing the search space. However, the complexity of eNBCH kernels with Viterbi algorithm is still high for practical applications.


\section{BACKGROUND}\label{window}

\textcolor{black}{In this section, after providing a brief background about the {\em channel polarization} of the polar codes constructed with large kernels and about {\em successive cancellation} (SC) decoder, we review the conventional reduced-complexity {\em window decoder} for decoding these kernels.}
\textcolor{black}{\subsection{Polar Codes}}

Consider a binary input discrete memoryless channel (B-DMC) $W : \mathcal{X} \rightarrow \mathcal{Y}$ with input alphabet $\mathcal{X} = \{0, 1\}$, output alphabet $\mathcal{Y}$, and transition probabilities $W(y|x)$,
where $x \in \mathcal{X}$, $y \in \mathcal{Y}$ and $W(y|x)$ is the conditional probability, the channel output $y$ given the transmitted input $x$. 
An $(N=l^n, k)$ polar code based on the $l \times l$ polarization kernel $K$ is a linear block code generated by $k$ rows of $G_n=K^{\otimes n}$,
and ${\otimes n}$ is $n$-times Kronecker product of matrix with itself. In order to polarize, none of the column permutations of the kernel $K$ should result in an upper triangular matrix.
Note that Arikan's polarization kernel, $F_2=\footnotesize \begin{pmatrix}
1 & 0\\
1 & 1
\end{pmatrix}$, is a special case of the polarization kernel $K$.

By encoding the binary input vector $u_0^{n-1}$ as $c_0^{n-1}=u_0^{n-1} G_n$, it was shown in \cite{Korada} that the transformation $G_n$ splits the B-DMC channel $W(y|x)$ into $N=l^n$ subchannels 
\begin{equation}
\begin{aligned}
W_{n, K}^{(i)}(u_0^i|y_0^{N-1})& =
\frac{W_{n, K}^{(i)}(y_0^{N-1},u_0^{i-1}|u_i)}{2 W(y_0^{N-1})}\\
& = \textcolor{black}{\sum_{u_{i+1}^{N-1} \in {\mathbb{F}_2}^{N-i-1}}} \prod _{i=0}^{N-1} W((u_0^{N-1}G_n)_i|y_i)\label{syntesis} 
\end{aligned}
\end{equation}
with capacities converging to $0$ or $1$ as $N \rightarrow \infty$, \textcolor{black}{where $\mathbb{F}_2$ is the binary field.}
Let's denote by $\mathcal{F}$ the set of the \textcolor{black}{indices} of subchannels with the lowest reliabilities. Then, setting $|\mathcal{F}|=N-k$ entries of the input vector $u_0^{N-1}$ to zero (frozen bits) and using the remaining entries to the information bits payload, will provide almost error-free communication.
 
 At the decoder side, the successive cancellation (SC) decoder first computes the $i$-th log likelihood ratio (LLR) in each step $i$, according to the following formula:
 \begin{equation}
 S_{n}^{(i)}(u_0^{i-1}, y_0^{N-1})\overset{\Delta}{=} \ln \frac{W_{n, K}^{(i)}(u_0^{i-1},u_i=0|y_0^{N-1})}{W_{n, K}^{(i)}(u_0^{i-1},u_i=1|y_0^{N-1})},\label{syntesis} 
\end{equation}
and then it sets the estimated bit $\hat{u}_i$ to the most likely value according to the following rule:
\begin{equation}
\hat{u}_i(\hat{u}_0^{i-1}, y_0^{N-1})=
\begin{cases}
u_i \   \ \textup{if} \ \ i \in \mathcal{F}\\
0,  \  \  \textup{if} \ \ i \in \mathcal{F}^C \ \& \  \textcolor{black}{S_{n}^{(i)}(u_0^{i-1}, y_0^{N-1})} \geq 0 \\ \label{uhat}
1, \   \ \textup{if} \ \ i \in \mathcal{F}^C \ \& \  \textcolor{black}{S_{n}^{(i)}(u_0^{i-1}, y_0^{N-1})} < 0. 
\end{cases}
\end{equation}

The complexity of this decoder for a polarization kernel $K$ of dimension $l \times l$ is $O(2^l N \log_l N)$. Methods for marginally reducing this complexity were proposed in \cite{Huang} and \cite{Land}; however, even for small $l$ this is still not practical.

\textcolor{black}{\subsection{Window Decoding}}
\textcolor{black}{ This method, introduced in \cite{TrifonovRS}, reduces the complexity of SC decoding without any performance loss. It achieves this goal by exploiting the relationship between the given kernel $K$ and Ar\i kan's kernel $(K_A=F_2^{\otimes t})$. }

\begin{figure}[h]
\centering
\includegraphics[width=0.6\linewidth]{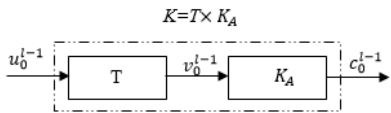}
\caption{Encoder block diagram}
\label{fig:WD}
\end{figure}

If we write the $l \times l$ $(n=1)$ polarization kernel $K$ with $l=2^t$ 
as a product of the Arikan's kernel with another matrix $T$, $K=T K_A$, then encoding is given by $c_0^{l-1}=v_0^{l-1} K_A$ and we have $c_0^{l-1}= u_0^{l-1}K$, where $u_0^{l-1}=v_0^{l-1} T^{-1}$ \textcolor{black}{(see Fig. \ref{fig:WD}).} 
Now, it is possible to reconstruct $u_0^i$ from $v_0^{\tau_i}$ where $\tau_i$ is the position of the last non-zero bit in the $i-$th row of $T^{-1}$. The relation between the vectors $v_0^{l-1}, u_0^{l-1}$ can be written as 
\begin{equation*}
    \theta ^{\prime} {(u_{l-1}, \ldots , u_1, u_0, v_0, v_1, \ldots, v_{l-1})}^{tr}=0,
\end{equation*}
where $\theta ^{\prime} =({S}|I)$ and ${S}$ is the $l \times l$ matrix obtained by transposing matrix  $T$ and reversing the order of the columns. Applying row operations can transform matrix $\theta ^{\prime} $ into a {\em minimum-span form} $\theta$, such that the $i$-th row starts in the $i$-th column, and ends in column $z_i$, where all $z_i$s are distinct. If we denote $j_i=z_{l-1-i}-l$ and $h_i=  \max_{0 \leq i^{\prime} \leq i} j_{i^{\prime}}$, one can express $u_i$ as
\begin{equation}
    u_{i}=\sum_{s=0}^{i-1} u_s \theta_{l-1-i,l-1-s} + \sum_{t=0}^{j_i} v_t \theta_{l-1-i, l+t}.\label{vu}
\end{equation}
As a result, the $i$-th bit channel of kernel $K$ in terms of the  $h_i$-th bit channel of kernel $K_A$ is


\begin{equation}
\begin{aligned}
    W_{1,K}^{(i)}(u_0^i|y_0^{l-1}) &= \sum_{v_{D_i} \in \{0, 1\}^{|D_i|}} W_{1,K_A}^{(h_i)}(v_0^{h_i}|y_0^{l-1})\\
    &= \sum_{v_{D_i} \in \{0, 1\}^{|D_i|}} \sum_{v_{h_i+1}^{l-1}} W_{1,K_A}^{(l-1)}(v_0^{l-1}|y_0^{l-1}), \label{LLR}
\end{aligned}
\end{equation}
where $D_i=\{0, 1, ..., h_i\} \setminus \{j_0, j_1, ..., j_i\}$ is the decoding window \textcolor{black}{and} $|D_i|=h_i-i$. Note that the set of the vectors $v_0^{h_i}$ in (\ref{LLR}) satisfies (\ref{vu}). 

In particular, (\ref{vu}) and (\ref{LLR}) imply that given the previous decoded bits $\hat{u}_0^{i-1}$, one can decode $u_i$ by using the SC decoder and calculating the transition probabilities $W_{1,K_A}^{(h_i)}(v_0^{h_i}|y_0^{l-1})$ for all $2^{|D_i|}$ values of $v_{D_i}$, corresponding to the input bits in $v_0^{h_i}$ that are not decoded yet. Then, for the evaluation of (\ref{syntesis}), one needs $2^{|D_i|+1}$ operations for both  ${W_{1, K}^{(i)}(u_0^{i-1},u_i|y_0^{N-1})}$ for $i$-th bit channel of the kernel $K$.
Thus, the $|D_i|$'s determine the window decoding complexity, \cite{TrifonovRS}.

\textcolor{black}{
Following \cite{Vletter}, (\ref{LLR}) is approximated as 
\begin{equation}
    \begin{aligned}
    \widetilde{W}_{1,K}^{(i)}(u_0^i|y_0^{l-1}) & = \max_{v_{D_i} \in \{0, 1\}^{|D_i|}} \widetilde{W}_{1,K_A}^{(h_i)}(v_0^{h_i}|y_0^{l-1})\\
    &= \max_{v_{D_i} \in \{0, 1\}^{|D_i|}} \max_{v_{h_i+1}^{l-1}} W_{1,K_A}^{(l-1)}(v_0^{l-1}|y_0^{l-1}). 
\end{aligned}
\end{equation}}

\textcolor{black}{As a result, the output LLRs $S_{1}^{(i)}, i=0, \ldots l-1$  in (\ref{syntesis}), can be approximated by:
\begin{equation}
\begin{aligned}
   & \widetilde{S}_1^{(i)}  = \ln \frac{\widetilde{W}_{1,K}^{(i)} (u_0^{i-1}, u_i=0|y_0^{l-1})}{\widetilde{W}_{1,K}^{(i)} (u_0^{i-1}, u_i=1|y_0^{l-1})}\\
    & = \max_{v_{D_i}\in\mathcal{Z}_{i,0}}\ln {\widetilde{W}}_{ 1,K_A}^{(h_i)}(v_0^{h_i}|y_0^{l-1})-\max_{v_{D_i} \in \mathcal{Z}_{i,1}}\ln {\widetilde{W}}_{1,K_A}^{(h_i)}(v_0^{h_i}|y_0^{l-1})\\
    &= \max_{v_{D_i} \in \mathcal{Z}_{i,0}} R(v_0^{h_i}|y_0^{l-1}) -\max_{v_{D_i}\in \mathcal{Z}_{i,1}} R(v_0^{h_i}|y_0^{l-1}),\label{S}
    \end{aligned}
\end{equation}
where $\mathcal{Z}_{i,b}=\{v_{D_i}|v_{D_i} \in \{0,1\}^{|D_i|}$, where $u_i=b\}$ and $R(v_0^{h_i}|y_0^{l-1})$ is the log-likelihood of a path $v_0^{h_i}$ and it can be obtained according to \cite{TrifonovScore} as
\begin{equation}
\begin{aligned}
    R(v_0^{h_i}|y_0^{l-1})&=\ln {\widetilde{W}}_{1,K_A}^{(h_i)}(v_0^{h_i}|y_0^{l-1})\\
    &=R(v_0^{h_i-1}|y_0^{l-1})+\tau (S_t^{(h_i)}(v_0^{h_i-1},y_0^{l-1}),v_{h_i}),
    \end{aligned} \label{R}
\end{equation}
where 
\begin{equation}
    \tau (S, v) = \begin{cases}
    0,  \ \ \ \ \ \  \text{sgn}(S)=(-1)^v\\
    -|S|, \ \ \text{otherwise}.
    \end{cases}
\end{equation}
is the penalty function\footnote{Note that $R( \cdot | y_0^{l-1})$ is initialized to $0$ for an empty sequence $v_0^{h_i}$.}  and $S_t^{(h_i)}(v_0^{h_i-1},y_0^{l-1})$ is the modified log-likelihood ratio defined as 
\begin{equation}
    S_t^{(h_i)}(v_0^{h_i-1},y_0^{l-1}) = \ln \frac{\widetilde{W}_{t, K_A}^{(h_i)} (v_0^{h_i-1}, v_{h_i}=0|y_0^{l-1})}{\widetilde{W}_{t, K_A}^{(h_i)} (v_0^{h_i-1}, v_{h_i}=1|y_0^{l-1})},
\end{equation}
The recursive expressions for the modified log-likelihood ratio of the $i$-th bit channel for $\lambda = 1, 2, \ldots, t$ can be obtained as
\begin{eqnarray}
S_{\lambda}^{(2i)}(v_0^{2i-1},y_0^{l-1})&=& Q(a,b),\\
S_{\lambda}^{(2i+1)}(v_0^{2i-1},y_0^{l-1})&=& P(a,b,v_{2i}),
\end{eqnarray}
where $a=S_{\lambda-1}^{(i)}(v_{0,e}^{2i-1} \oplus v_{0,o}^{2i-1}, y_{0,e}^{l-1})$ , $b=S_{\lambda-1}^{(i)}(v_{0,o}^{2i-1}, y_{0,e}^{l-1})$, $Q(a,b)= \text{sgn}(a) \text{sgn} (b) \min(|a|,|b|)$, $P(a,b,c)=(-1)^c a + b$ and $m=2^\lambda$. Note that these expressions are the same as the min-sum approximation of the list SC algorithm \cite{Tal}. }

\textcolor{black}{In this paper, we use the approximated LLR formula in (\ref{S})
to implement the SC decoder. \textcolor{black}{The approximate complexity (AC) of the LLR domain} 
implementation of the window decoder for the kernel $K$ of size $l$ is} 
\begin{eqnarray}
\psi(K)\!\!&\!\!=\!\!&\sum_{i=0}^{l-1} \phi(i) \text{~~~where}\label{Cost}\\
\phi(i)\!\!&\!\!=\!\!&\!\!
    \begin{cases}
        2^{|D_i|+1} -1 +\Lambda(i) \ \ \ \text{if} \ \ h_i>h_{i-1}  \& |D_i|>0\\ 
       
       C_i  \ \ \ \ \ \ \ \ \ \ \ \ \ \ \  \ \ \ \ \ \ \ \text{if} \ \ h_i>h_{i-1}  \& |D_i|=0\\
       1 \ \ \ \ \ \ \ \ \ \ \ \ \ \ \ \ \ \ \ \ \ \ \ \ \text{if} \ \ h_i=h_{i-1},
    \end{cases} \nonumber
\end{eqnarray}
with $\Lambda(i)=\sum_{h=h_{i-1}+1}^{h_i} 2^{(h+B(h)-i)}$, $B(h)=\log_2{(C+1)}$ where $C$ is the computational cost of a bit channel of $K_A$, \cite{T32}. For the $i$-th bit channel, let $s$ be the largest integer such that $2^s$ divides $i$, then $C= C_i=2^{s+1}-1$ and $C_0=2^t-1$.

\section{Proposed Algorithm}


\textcolor{black}{In this section, we propose an algorithm to modify 
large kernels to reduce the complexity of the window decoding.}
First, we explain the motivation and  the intuition behind our method. Then, we present the algorithm which achieves this goal.

As it is stated in Section {\ref{window}}, the size of the window, $|D_i|=h_i-i$, determines the complexity of the window decoder, when $h_i>h_{i-1}$. Tables \ref {complexity16} and \ref {complexity32} show $|D_i|$s for each of the kernels $K_{\text{eNBCH}}$, $K_\text{L}$ and $K_{\text{F}}$ constructed in \cite{Vera}, \cite{Lin} and \cite{Fazeli}, respectively. These tables show that the size of window for some of the bit channels is very large. \textcolor{black}{One solution  to this problem is to modify the kernel $K$ to reduce the size of the window, without altering the polarization properties of the kernel. Column permutation of the kernel $K$ \textcolor{black}{does not} affect its polarization properties, \cite{Fazeli}. However, finding the best permutation by exploring all $l!$ cases is not practical and a method for reducing the size of the search space is needed.}
On the other hand, there are many permuted kernels that have the same $h_i$, resulting in the same complexity.  However, reducing the search space to the permutations that result in the unique $h_i$'s is not trivial. One suboptimal method is to limit our search to the permuted kernels that result in matrix $T$ with as many rows with only one non-zero element. 
This implies that in the new column permuted kernel $K^{\prime}$, we want to have as many rows from $K_A$ as possible.
A threshold, $\textcolor{black}{M_{t}}$, to satisfy this rule will reduce the search space significantly. \textcolor{black}{$M_t$ is the maximum possible rows that the kernel $K$ can have from $K_A$.} After finding the permutations which satisfy this rule, we choose the ones which have the minimum complexities based on (\ref{Cost}). 
\textcolor{black}{Although this solution is sub-optimal, it still reduces the search space to a manageable size and finds a permutation that significantly reduces the complexity of window decoding. We refer to the  permutations given by the proposed algorithm as {\em good} column permutations.}

\begin{algorithm}
\small
\footnotesize
\DontPrintSemicolon
\SetAlgoLined
\SetNoFillComment
\SetKwInOut{Input}{input}
\SetKwInOut{Output}{output}
\SetKwRepeat{Repeat}{do}{while} 


\Input{ Kernel $K$ of size $l$ and  threshold $\textcolor{black}{M_{t}}$}
\Output{ Good column permutation $\pi$ and permuted kernel $K^{\prime}$}

Define Lists: $\text{C}, \text{R}, \text{M}$, $\text{TmpC}, \text{TmpR}, \text{TmpM}$\\

 \textcolor{black}{[$\text{C}$, $\text{R}$, $\text{M}$] $\gets [\{ \{\}\},\{\{1, 2, ..., l\}\},\{l\}]$;\tcp*{Init.}}

\textcolor{black}{[TmpC, TmpR, TmpM]$\gets [\{\},\{\},\{\}]$\tcp*{Init.}}
\While{{$(\text{\em C}=\{\{\}\})$}}{


    \For {$i \gets 1$ \KwTo $l$ }{\tcp*{find best cand.~$i$-th col.}
        
                \ForEach{\textcolor{black}{$[\kappa, \textcolor{black}{\rho}, \mu] \in [\text{\em C}, \text{\em R, M}$\em]}}{\tcp*{best cand. of $(i-1)$-th col.} 
                
            $\text{CandCol} \gets  \{1, 2, ..., l\} \setminus {[\kappa]};$ \tcp*{Cand. for $i$-th col.}

            \For {$m \gets 0$ \KwTo $\text{\em length}(\text{\em CandCol})-1$}{\tcp*{check all cand. for $i$-th col.}
                
                $[$\text{Cel}$, $\text{Rel}$,  $\text{Metric}$] \gets \text{CalculateMetric} ( i, {\text{CandCol}}, \kappa, m, \textcolor{black}{\rho}, \mu)$\\

                \If {$\text{\em Metric} \geq \textcolor{black}{M_{t}} $}{
                   
                     {PUSH} \textcolor{black}{$([\text{Cel}$, $\text{Rel}$, $\text{Metric}]$, $[\text{TmpC}$, $\text{TmpR}$, $\text{TmpM}])$;}\\ 
 }
             
            }
           
        }
        
          \eIf{\textcolor{black}{$([\text{\em TmpC}, \text{\em TmpR}, \text{\em TmpM}]= [\{\}, \{\}, \{\}])$} }{
          \textcolor{black}{$[\text{C}, \text{R}, \text{M}] \gets [\{\{\}\},\{\{1, 2, \ldots, l\}\},\{l\}]$;}\\
         \textbf{goto line 23}; \\
          }{
        \textcolor{black}{$[\text{C}, \text{R}, \text{M}] \gets    [\text{TmpC}, \text{TmpR}, \text{TmpM}]$;}\\
         \textcolor{black}{$[\text{TmpC}, \text{TmpR}, \text{TmpM}] \gets [\{\},\{\},\{\}]$;}
        }
    }
    $\textcolor{black}{\textcolor{black}{M_{t}}} \gets \textcolor{black}{M_{t}}-1$;\\
}
 
 $\pi=\argmin_{\kappa \in \text{C}} \Psi(\kappa(K)) $\\
\KwRet $K^{\prime}=\pi(K)$, $\pi$\\

     \SetKwFunction{Fce}{\text{CalculateMetric}}
   \SetKwProg{Fn}{subroutine}{:}{}
   \Fn{\Fce{$i,{\text{\em CandCol}}$, $\kappa$, ${m}$, $\textcolor{black}{\rho}$, ${\mu}$}}{
       
                $\text{Cel} \gets (\kappa, \text{CandCol}[m])$\tcp*{append 
                $\text{CandCol}[m]$ to $\kappa$}

                $SK \gets \{K[\textcolor{black}{\rho}][\text{Cel}]\}$\tcp*{List of rows of the subkernel $K[\textcolor{black}{\rho}][\text{Cel}]$ (with repetitions)}
                $SK_{A} \Leftarrow \{K_A[:][1:i]\}$\tcp*{Set of rows of the subkernel $K_{A}[:][1:i]$ (no repetitions)}
                $ \text{Metric} \gets \text{Calculate Metric with (\ref{metric})}$ for $SK_A$ and $SK$\\
            $\text{cnt} \gets 0$; \tcp*{counter}
                   \For {$ j \gets 0$ \KwTo $\mu-1$}{
                 \If{$SK[j] \in SK_{A}$}{
                $\text{cnt} \gets \text{cnt}+1$\\ $\text{Rel}[\text{cnt}] \gets \textcolor{black}{\rho}[j]$
                 }
                    
                }

       \KwRet $\text{Cel}$, $\text{Rel}$, $\text{Metric}$;
   }

\caption{Finding good column permutation}
\end{algorithm}

Algorithm 1 shows the process of finding good column permutations for an arbitrary kernel $K$. The inputs to this algorithm are the  kernel $K$ of size $l$ and a predefined threshold $\textcolor{black}{M_{t}}$. The outputs are good permutations $\pi$ and the resulting permuted kernel $K^{\prime}=\pi(K)$. In each step, the best candidates for the first $(i-1)$ columns of the {\em partially permuted kernel}  are used to examine all possible candidates for the $i$-th column to determine the best ones based on the following metric,
\begin{equation}
     \text{Metric}=\sum_{j \in SK} \mathbbm{1}_{SK_A}{(j)}, \label {metric}
\end{equation}
where $SK$ and $SK_A$ are the sets of some rows of the first $i$ columns of the partially permuted kernel and of $K_A$,
respectively, and $\mathbbm{1}$ is the indicator function. \textcolor{black}{In other words, for each possible candidate for the $i$-th column, the variable `Metric' counts the number of the rows of the partially permuted kernel that can be found in the first $i$ columns of $K_A$.}

Let's define $\text{C}$ as the list of the best candidates for the first $(i-1)$ columns, $\text{R}$ as the list of the  \textcolor{black}{indices} of the rows of the first $(i-1)$ columns of the partially permuted kernel, which belong to the set of the rows of the first $(i-1)$ columns of the kernel $K_A$. Let   $\text{M}$ be the list of the $\text{Metric}$s of the best candidates for the $(i-1)$ columns.
Note that for $i=1$, we assume that $\text{C} = \{\{\}\}$, $\text{R}=\{\{1, 2, ..., l\}\}$ and $\text{M}=\{l\}$.

The proposed algorithm finds the best candidates for the $i$-th column of the partially permuted kernel with the following steps: 
\begin{enumerate}
\item For each \textcolor{black}{$[\kappa, \textcolor{black}{\rho}, \mu]\in [\text{C}, \text{R}, \text{M}]$}, it determines the non yet selected columns as the  candidates $\text{CandCol}$ for the $i$-th column (line $8$). 
Then, for each of these  candidates, it follows the four steps:  
\begin{itemize}
\item Appends a candidate from $\text{CandCol}$ to $\kappa$ to obtain $\text{Cel}$, the list of the first $i$ possible columns. 
\item Picks each row of the sub-kernel $K[\textcolor{black}{\rho}][\text{Cel}]$ and puts it in the list $SK$ accounting for any repetitions and picks each row of the sub-kernel $K_A[:][1:i]$ and puts it in the {\em set} $SK_{A}$ (without repetitions). 
    
\item Counts the elements of $SK$  belonging to the set $SK_{A}$ and stores this number in $\text{Metric}$. Stores the corresponding \textcolor{black}{indices} of the rows belonging to set $SK_{A}$ in $\text{Rel}$.

\item Compares $\text{Metric}$ with the threshold $\textcolor{black}{M_{t}}$. If $\text{Metric} \geq \textcolor{black}{M_{t}}$, \textcolor{black}{it appends $\text{Cel}$ to the temporary list $\text{TmpC}$, $\text{Rel}$ to $\text{TmpR}$ and $\text{Metric}$ to $\text{TmpM}$, respectively.} 
\end{itemize}
\item If there is at least one candidate from $\text{CandCol}$ with $\text{Metric}\geq \textcolor{black}{M_{t}}$, it copies all the collected parameters of these candidates to \textcolor{black}{$[\text{C}$, $\text{R}$, $\text{M}]$} to use them for the next column selection; otherwise, it reduces the threshold $\textcolor{black}{M_{t}}$ and repeats the process from the beginning, with $i=1$.
\end{enumerate}

Finally, the algorithm continues this process until it finds the best candidates for all $l$ columns. Then, it outputs the good column permutations, among the candidates in $\text{C}$, which minimize the approximate complexity in (\ref{Cost}).


The algorithm significantly reduces the $l!$ search space to the candidates in list C by using the threshold $\textcolor{black}{M_{t}}$ and the $\text{Metric}$ (\ref{metric}). Then, it  finds the best ones among them.

\textcolor{black}{To find the initial value of the threshold $\textcolor{black}{M_{t}}$, we define two multisets, $\text{HWK}$ and $\text{HWA}$, containing the Hamming weights of the rows of $K$ and $K_A$, respectively.}
Then, $\textcolor{black}{M_{t}}$ which is the maximum possible threshold will be $\textcolor{black}{M_{t}}= |\text{HWK} \cap \text{HWA}|$.

\textcolor{black}{Here, we provide an example to illustrate how Algorithm 1 works.}

\begin{figure}[h]
\centering
\hspace{-4mm}
\includegraphics[width=0.8\linewidth]{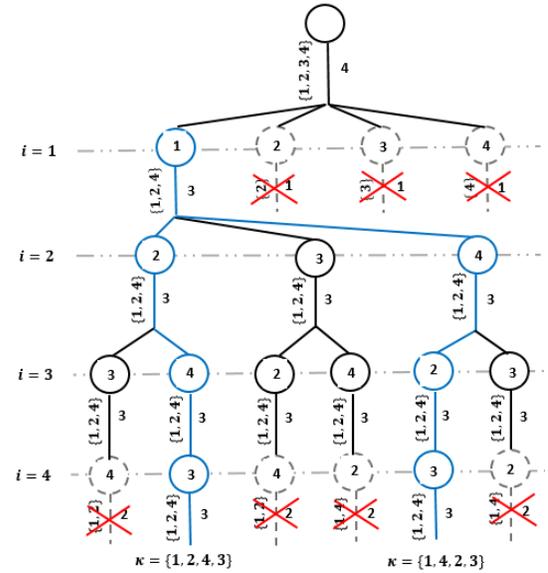}
\caption{\textcolor{black}{Example of finding good column permutation for kernel of size $4$.}}
\label{fig:example}
\end{figure}

\textcolor{black}{\textit{Example.} Consider $K=\footnotesize \begin{pmatrix}
1 & 0 & 0 & 0\\
1 & 1 & 0 & 0\\
0 & 0 & 1 & 0\\
1 & 0 & 0 & 1
\end{pmatrix}$. Since $\text{HWK}=\{1, 2, 1, 2\}$, and $\text{HWA}=\{1, 2, 2, 4\}$, then $M_t=3$. The initial values for the column candidates, row indices and Metrics are $C=\{\{\}\}, R=\{\{1, 2, 3, 4\}\}$ and $M=\{4\}$, respectively. Fig. \ref{fig:example} shows the steps of the algorithm. The values inside the circles are candidates for the $i$-th column (CandCol) and the values on the left and on the right hand-side of the branches are 'Rel' and 'Metric', respectively.}

\textcolor{black}{For each $i$, the algorithm checks all the candidates to find the ones which satisfy the condition $\text{Metric} \geq M_t$. For this example, two candidates $\{1, 2, 4, 3\}$ and $\{1, 4, 2, 3\}$  (blue paths) satify this condition. 
The algorithm reduces the search space from $4!$ to $2$ candidates. Then, it chooses the one which minimizes complexity (\ref{Cost}), as the good column permutation.}


We apply our algorithm to two kernels $K_{\text{F}}$ and $K_\text{L}$ of sizes $16$ constructed in \cite{Fazeli} and \cite{Lin}, respectively, and to the kernels $K_{\text{eNBCH}}$ of sizes $16$ and $32$, \cite{Vera}. The error exponents (EE), scaling exponents (SE) and good permutations resulting from our algorithm for these kernels are given in TABLE \ref{permutation}. Due to space limitations, for $K_{\text{eNBCH}}$ of size $32$, we wrote only one of the good permutations.

Although the proposed algorithm is sub-optimal, we conjecture that the obtained permutations for $K_{\text{eNBCH}}$ may be optimal. 



\begin{table}[h]
\begin{center}
\scalebox{0.7}{
\begin{tabular}{|P{0.5cm}|P{1cm}|P{0.95cm}|P{0.84cm}|P{6.4cm}|}
\hline
 Size & Kernel & EE & SE & Permutation \\ 
\hline
 \multirow{10}{*}{16} & \multirow{1}{*}{$K_{\text{eNBCH}}$} & $0.51828$ & $3.3957$ & \multirow{1}{*}{$1, 3, 4, 7, 6, 2, 12, 10, 5, 11, 8, 13, 9, 16, 14, 15$} \\  
\cline{2-5}
 &  \multirow{8}{*}{$K_{\text{L}}$} & \multirow{8}{*}{$0.51828$}  &
 \multirow{8}{*}{$3.3627$}  &
            $1, 4, 3, 7, 2, 5, 6, 12, 14, 15, 9, 8, 11, 13, 10, 16$\\ 
           & & & & $1, 4, 3, 7, 2, 5, 6, 16, 14, 15, 9, 8, 11, 13, 10, 12$\\ 
           & & & & $1, 4, 3, 8, 2, 5, 6, 12, 14, 15, 9, 7, 11, 13, 10, 16$\\ 
           & & & & $1, 4, 3, 8, 2, 5, 6, 16, 14, 15, 9, 7, 11, 13, 10, 12$\\ 
           & & & & $1, 4, 3, 12, 2, 5, 6, 7, 14, 15, 9, 16, 11, 13, 10, 8$\\ 
           & & & & $1, 4, 3, 12, 2, 5, 6, 8, 14, 15, 9, 16, 11, 13, 10, 7$\\ 
           & & & & $1, 4, 3, 16, 2, 5, 6, 7, 14, 15, 9, 12, 11, 13, 10, 8$\\ 
           & & & & $1, 4, 3, 16, 2, 5, 6, 8, 14, 15, 9, 12, 11, 13, 10, 7$\\
           
\cline{2-5}
 & \multirow{1}{*}{$K_{\text{F}}$} & $0.51828$ & $3.356$ & \multirow{1}{*}{$16, 12, 14, 10, 8, 4, 6, 2, 15, 11, 13, 9, 7, 3, 5, 1$}\\
\hline
\multirow{2}{*}{32} & \multirow{2}{*}{$K_{\text{eNBCH}}$} & \multirow{2}{*}{$0.537$} & \multirow{2}{*}{$3.1221$} & $1, 2, 3, 20, 4, 7, 21, 13, 5, 31, 8, 29, 22,$\\
& & & & $10, 14, 25, 6, 12, 32, 19, 9, 24, 30, 28,$\\
& & & & $23, 27, 11, 18, 15, 16, 26, 17$\\
\hline
\end{tabular}%
}
\end{center}
\vspace{-6mm}
\caption{Good permutations found by Algorithm I for different kernels.}
\label{permutation}
\end{table}

\section{Analysis and Simulation Results}

\textcolor{black}{In this section, we first analyze the computational complexity of the kernels and compare the complexity of them before and after applying the permutations. Then, we compare their performances and also their real-time complexities with each other and also with the Ar\i kan's kernel under SC and SCL decoders.} 
\textcolor{black}{Note that for the computational complexity, we use arithmetical complexity as the number of summation and comparison operations to calculate the LLR in (\ref{S}).}

 Tables \ref{complexity16} and  \ref{complexity32} show $h_i$ and $|D_i|$ for each $i \in \{0, 1, \ldots l-1\}$ of the different kernels of sizes $16$ and $32$ before and after applying the column permutations. The corresponding approximate  computational complexities  $(AC)$ using the expression (\ref{Cost}) as well as the computational complexities $(CC)$ using CSE  algorithm proposed in \cite{T16} are also given in these tables. 
 
 
 
 
 It can be observed that the good column permutations obtained from our algorithm can reduce the maximum size of the window from $12$ to $4$ for $K_{\text{eNBCH}}$ of size $16$, from $7$ to $4$ for $K_{\text{F}}$, from $12$ to $5$ for $K_{\text{L}}$ and from $28$ to $17$ for $K_{\text{eNBCH}}$ of size $32$. As a result, the computational complexity $CC$ of the window decoder after applying the column permutations is reduced from $38089$ to $465$ for $K_{\text{eNBCH}}$ of size $16$,  from $1851$ to $517$ for $K_{\text{F}}$ and from $38089$ to $728$ for $K_\text{L}$ without any performance loss. 
 Also, for $K_{\text{eNBCH}}$ of size $32$, the good column permutation results in complexity reduction by factor of $1192$, as compared to the original kernel. 
 
Note that after applying column permutation proposed in \cite {TNew} on $K_{\text{eNBCH}}$ of size $16$, Viterbi decoder needs $5019$ operations, while using our best permutations for window decoding requires only $446$.
Indeed, using window decoder for the kernels constructed in \cite{TNew} requires $3000$ and $6714318$ operations approximately, while applying \textcolor{black}{Viterbi} decoder needs $5019$ and $299235$ operations, respectively. 
\textcolor{black}{Moreover, window decoding with the obtained kernels provides much more efficient implementation of kernel processing compared to the algorithm presented in \cite{Huang}. For example, the algorithm proposed in \cite{Huang} requires $16 \times 2487 \times 15 = 596880$
operations for the $16 \times 16$ kernel $K^{\prime}_L$, while window decoding using the CSE algorithm, for the same kernel, requires only $728$, on average.}


\begin{table*}
\begin{center}
\scalebox{0.72}{
\begin{tabular}{|P{0.2cm}||P{0.25cm}|P{0.45cm}|P{0.8cm}|P{0.8cm}|P{0.25cm}|P{0.45cm}|P{0.55cm}|P{0.55cm}||P{0.25cm}|P{0.45cm}|P{0.55cm}|P{0.55cm}|P{0.25cm}|P{0.45cm}|P{0.55cm}|P{0.55cm}||P{0.25cm}|P{0.45cm}|P{0.8cm}|P{0.8cm}|P{0.25cm}|P{0.45cm}|P{0.55cm}|P{0.55cm}|}
\hline

\multirow{3}{*}{$i$} & \multicolumn{4}{|P{2cm}|}{{$K_{\text{eNBCH}}$}} &
\multicolumn{4}{|P{2cm}||}{{$K^{\prime}_{\text{eNBCH}}$}}&
\multicolumn{4}{|P{2cm}|}{{\textcolor{black}{$K_{\text{F}}$}}} &
\multicolumn{4}{|P{2cm}||}{{\textcolor{black}{$K^{\prime}_{\text{F}}$}}} &
\multicolumn{4}{|P{2cm}|}{{$K_{\text{L}}$}} &
\multicolumn{4}{|P{2cm}|}{{$K^{\prime}_{\text{L}}$}} \\


\cline{2-25}
  & $h_i$ & $|D_i|$ & $AC_i$ & $CC_i$ & $h_i$ & $|D_i|$ & $AC_i$ & $CC_i$ & $h_i$ & $|D_i|$ & $AC_i$ & $CC_i$ & $h_i$  & $|D_i|$ & $AC_i$ & $CC_i$ & $h_i$ & $|D_i|$ & $AC_i$ & $CC_i$ & $h_i$ & $|D_i|$ & $AC_i$ & $CC_i$ \\ 
\hline
 $0$ & $0$ & $0$ & $15$ & $15$ & $0$  & $0$ &  $15$  & $15$ & $0$ & $0$ & $15$ & $15$ & $0$  &$0$ & $15$ & $15$ & $0$ & $0$ & $15$ & $15$  & $0$  & $0$ & $ 15$  & $15$\\ 
\hline
  $1$ & $13$ &$12$& $39793$ & $17326$ & $4$   & $3$ & $97$ & $63$ & $8$ & $7$ & $2673$ & $972$ & $4$  &$3$ & $97$ & $63$ & $13$ &$12$& $39793$ & $17326$ & $4$ & $3$ & $97$ & $63$\\   
\hline
  $2$ & $14$ & $12$ & $24575$ & $20735$ & $4$   & $2$ & $1$ & $1$ & $8$ & $6$ &$ 1$ & $1$ & $4$  & $2$ & $1$ & $1$ & $14$ & $12$ & $24575$ & $20735$ & $4$  & $2$ & $1$ & $1$\\ 
\hline
  $3$ & $14$ & $11$ & $1$ & $1$ & $4$   & $1$ &  $1$ & $1$ & $8$ & $5$& $1$ & 1 & 4   & $1$ &  $1$ & $1$ & $14$ & $11$ & $1$ & $1$ & $4$ & $1$ & $1$  & $1$\\ 
\hline
 $4$ & $14$ & $10$ & $1$ & $1$ & $8$  & $4$ & $323$ & $ 127$ & $8$ &$4$ & $1$ & $1$ & $8$ & $4$ & $323$ & $127$ & $14$ & $10$ & $1$ & $1$  & $8$  & $4$ & $323$ & $127$\\ 
\hline
 $5$ & $14$ & $9$ & $1$ & $1$ &  $9$   & $4$ & $63$ & $48$ & $10$ & $5$ & $223$ & $223$ & $9$ &$4$ & $63$ & $48$ & $14$ & $9$ & $1$ & $1$ & $10$ & $5$ & $223$ & $207$\\ 
\hline
  $6$ & $14$ & $8$ & $1$ & $1$ &  $9$   & $3$ &  $1$ & $1$ & $12$ & $6$ & $703$ & $575$ & $10$ &$4$ & $95$ & $95$ & $14$ & $8$ & $1$ &$1$ & $10$ & $4$ & $1$  & $1$ \\ 
\hline
 $7$ & $14$ & $7$ & $1$ & $1$ & $10$   & $3$ & $47$ & $47$ & $12$ & $5$ & $1$ & $1$ & $10$   &$3$ & $1$ & $1$ & $14$ & $7$ & $1$ & $1$ & $12$ & $5$ & $351$ & $279$\\
\hline
  $8$ & $14$ & $6$ & $1$ & $1$ &  $12$   & $4$ & $175$ & $143$ & $12$ &$4$ & $1$ & $1$ & $12$ &$4$ & $175$  & $143$ & $14$ & $6$ & $1$ & $1$ & $12$ & $4$ & $1$ & $1$\\   
\hline
  $9$ & $14$ & $5$ & $1$ & $ 1$ & $ 12$   & $3$ & $1$ & $1$ & $12$ &$3$ & $1$ & $ 1$ & $12$   &$3$ & $1$ & $1$ & $14$ & $5$ & $1$ & $1$ & $12$  & $3$ & $1$ & $1$\\ 
\hline
 $10$ & $14$ & $4$ & $1$ & $1$ & $12$   & $2$ & $1$ & $1$ & $12$ & $2$ & $1$ & $1$ & $12$ &$2$ & 1 & $1$ & $14$ & $4$ & $1$ & $1$  & $12$  & $2$ & $1$  & $1$\\ 
\hline
  $11$ & $14$ & $3$ & $1$ & $1$ &  $12$   & $1$ & $1$ & $1$ & $14$ & $3$ & $55$ & $55$ & $12$  & $1$& $1$ & $1$ & $14$ & $3$ & $1$ & $1$ & $13$  & $2$ & $15$ & $15$\\  
\hline
  $12$ & $14$ & $2$ & $1$ & $1$ &  $12$  & $0$ & $1$ & $1$ & $14$ &$2$ & $1$ & $1$ & $13$  &$1$ & $7$ & $7$ & $14$ & $2$ & $1$ & $1$  & $13$ & $1$ & $1$ & $1$\\ 
\hline
 $13 $ & $14$ & $1$ & $1$ & $1$ & $14$   & $1$ & $13$ & $13$ & $14$ & $1$& $1$ & $1$ & $14$ &$1$ & $11$ & $11$ & $14$ & $1$ & $1$ & $1$ & $14$ & $1$ & $11$ & $11$\\ 
\hline
  $14$ & $14$ & $0$ & $1$ & $1$ &  $14$  & $0$ & $1$ & $1$ & $14$ & $0$ & $1$ & $1$ & $14$ &$0$ & $1$ & $1$ & $14$ & $0$ & $1$ & $1$ & $14$ & $0$ & $3$ &$3$ \\   
\hline
  $15$ & $15$ & $0$ & $1 $ & $1$ &  $15$   & $0$ & $1$ & $1$ & $15$ &$0$ & $1$ & $1$ & $15$ &$0$ & $1$ & $1$ & $15$ & $0$ & $1$ & $1$ & $15$ & $0$ & $1$ & $1$\\ 
\hline
\multicolumn{1}{|P{0.15cm}||}{} &
\multicolumn{4}{|P{3cm}|}{$CC=38089$} &
\multicolumn{4}{|P{2.5cm}||}{{ $ \bf {CC=465}$}} &
\multicolumn{4}{|P{2cm}|}{$CC=1851$} &
\multicolumn{4}{|P{2.5cm}||}{$\bf {CC=517}$} &
\multicolumn{4}{|P{3cm}|}{$CC=38089$} &

\multicolumn{4}{|P{2cm}|}{$\bf {CC=728}$} \\
\hline
\multicolumn{17}{P{5.1cm}}{}\\
\end{tabular}
}
\end{center}
\vspace{-6mm}
\caption{Comparison of different size $16$ kernels.}
\label{complexity16}
\end{table*}

\begin{table}
\begin{center}
\scalebox{0.65}{
\begin{tabular}{|P{0.3cm}|P{0.3cm}|P{0.5cm}|P{1.7cm}|p{0.3cm}|P{0.3cm}|p{0.5cm}|P{1.2cm}|}
\hline

\multicolumn{4}{|P{2.5cm}|}{{$K_{\text{eNBCH}}$}} &
\multicolumn{4}{|P{2.5cm}|}{{$K^{\prime}_{\text{eNBCH}}$}}\\

\cline{1-8}
   $i$ & $h_i$ & $|D_i|$ & $AC_i$ &  $i$ &$h_i$ & $|D_i|$ & $AC_i$  \\ 
\hline
  $0$ & $0$ &$0$ &$31$  & $0$ & $0$& $0$ &$31$  \\ 
\hline
 $1$ & $29$ &$28$ & $2.6e+9$ & $1$ &$1$ & $0$& $3$\\   
\hline
   $2$&$29$ & $27$& $1$  & $2$& $2$&$0$&$5$ \\ 
\hline
   $3$ &$30$ &$27$& $805306367$&$3$& $4$ & $1$& $21$\\ 
\hline
  $4$& $30$&$26$ &$1$ & $4$&$8$ & $4$& $323$\\ 
\hline
  $5$& $30$& $25$& $1$& $5$& $16$& $11$&$75551$ \\ 
\hline
  $6$&$30$ & $24$& $1$ & $6$&$16$ &$10$ &$1$\\ 
\hline
  $7$&$30$ & $23$ & $1$ & $7$& $24$&$17$&$2738175$ \\
\hline
   $8$& $30$& $22$&$1$ & $8$&$24$ &$16$ &$1$\\   
\hline
   $9$&$30$ & $21$& $1$&  $9$ & $24$& $15$&$1$\\ 
\hline
  $10$&$30$ & $20$ & $1$& $10$&$24$ &$14$&$1$ \\ 
\hline
   $11$&$30$ & $19$ & $1$& $11$ &$24$ & $13$& $1$\\  
\hline
   $12$& $30$& $18$& $1$& $12$ &$24$ &$12$&  $1$\\ 
\hline
  $13$&$30$ & $17$& $1$& $13$&$24$ &$11$ &$1$\\ 
\hline
   $14$&$30$ & $16$ &$1$ & $14$&$24$ & $10$& $1$\\   
\hline
    $15$&$30$ & $15$& $1$& $15$& $24$&$9$ &$1$\\

\hline
  $16$ &$30$ & $14$&$1$ & $16$&$28$ &$12$ &$50175$ \\ 
\hline
 $17$ & $30$& $13$ & $1$ &$17$ &$28$& $11$&$1$ \\   
\hline
    $18$&$30$ & $12$&$1$ &  $18$&$28$ &$10$& $1$\\ 
\hline
   $19$ &$30$ & $11$ & $1$&   $19$& $28$& $9$&$1$\\ 
\hline
  $20$ & $30$& $10$ & $1$&  $20$& $28$&$8$ &$1$\\ 
\hline
  $21$ &$30$ & $9$ &$1$ & $21$ &$28$& $7$&$1$\\ 
\hline
  $22$ & $30$& $8$ &$1$ & $22$ &$28$& $6$&$1$\\ 
\hline
  $23$ & $30$& $7$& $1$ & $23$&$28$ & $5$&$1$\\
\hline
   $24$& $30$ &$6$ & $1$&  $24$&$28$ & $4$ & $1$\\   
\hline
   $25$&$30$ &$5$& $1$& $25$& $28$& $3$ &$1$\\ 
\hline
  $26$& $30$& $4$& $1$& $26$&$30$ &$4$& $111$\\ 
\hline
   $27$&$30$ & $3$& $1$&  $27$& $30$& $3$&$1$\\  
\hline
   $28$&$30$ & $2$& $1$& $28$&$30$ &$2$& $1$\\ 
\hline
  $29$& $30$& $1$ &$1$& $29$&$30$ &$1$ &$1$\\ 
\hline
    $30$ &$30$ & $0$&$1$ & $30$& $30$& $0$&$1$\\   
\hline
   $31$& $31$& $0$& $3$& $31$ &$31$& $0$ &$3$\\ 
\hline
\multicolumn{4}{|P{3.5cm}|}{$AC=3.4144e+09$} &
\multicolumn{4}{|P{3.5 cm}|}{$\bf{AC=2864420}$} \\
\hline


\end{tabular}
}

  \caption{Comparison of size $32$ eNBCH kernels.}
\label{complexity32}

\end{center}
\vspace{-8mm}
\end{table}

\begin{figure}[h]
\vspace{-5mm}
\includegraphics[width=1.0\linewidth]{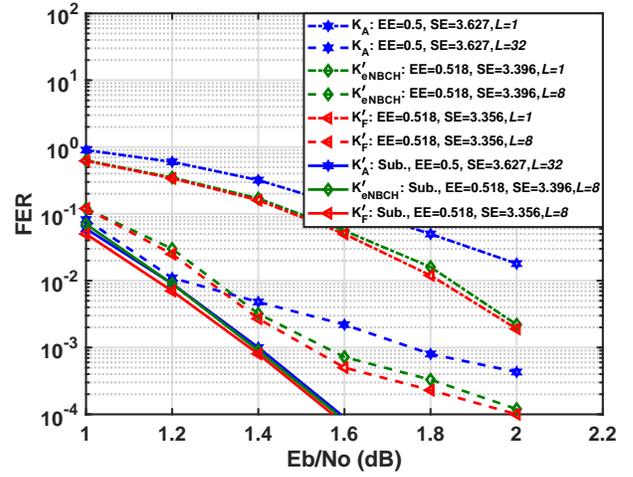}
\caption{Performance comparison for polar (sub)codes with $N=4096$ and $R=0.5$, for different kernels.}
\label{fig:Performance}
\end{figure}

\begin{figure}
     \centering
     \begin{subfigure}[b]{0.38\textwidth}
         \centering
         \includegraphics[width=\textwidth]{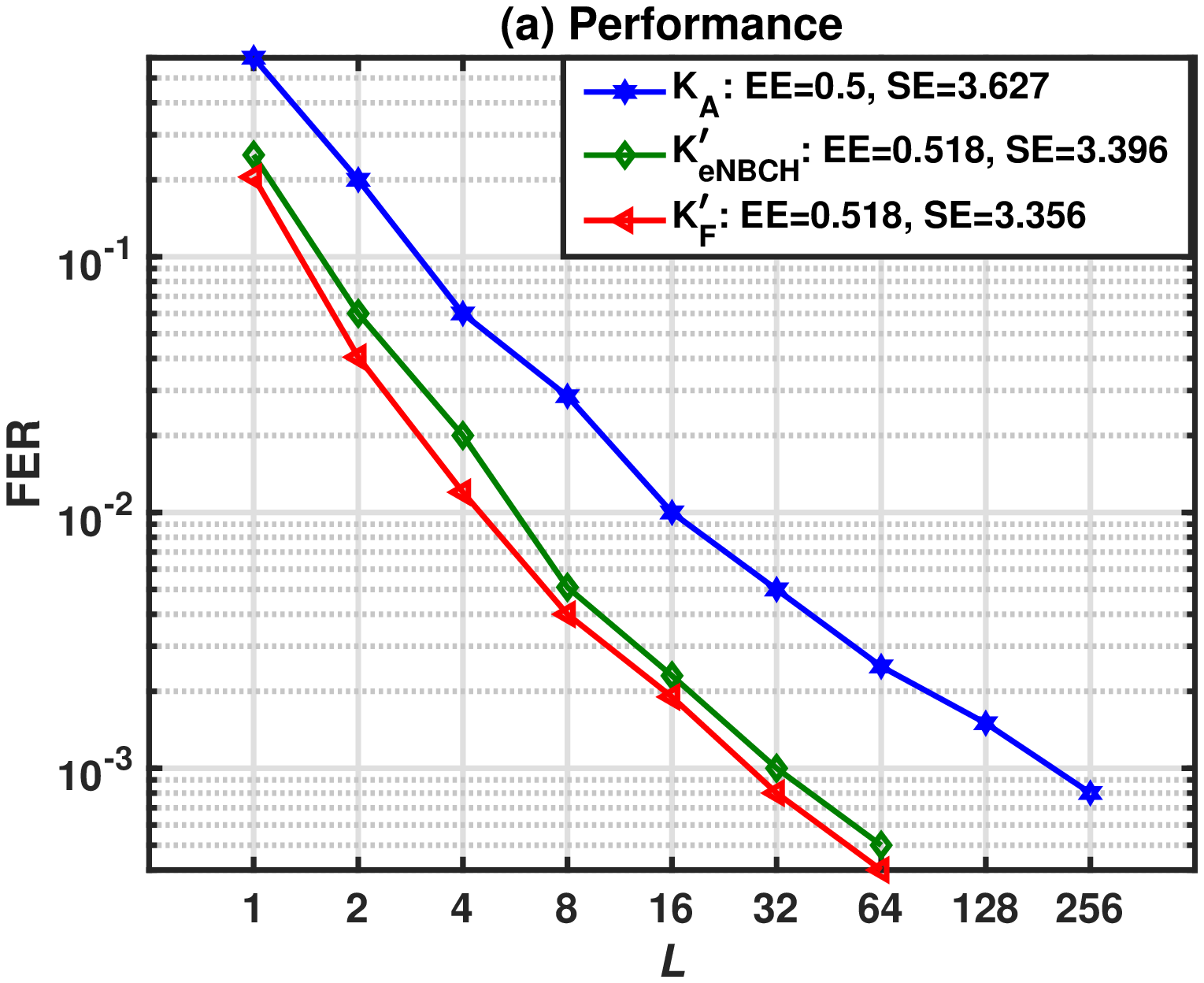}
         
     \end{subfigure}
     \hfill
     \begin{subfigure}[b]{0.38\textwidth}
         \centering
         \includegraphics[width=\textwidth]{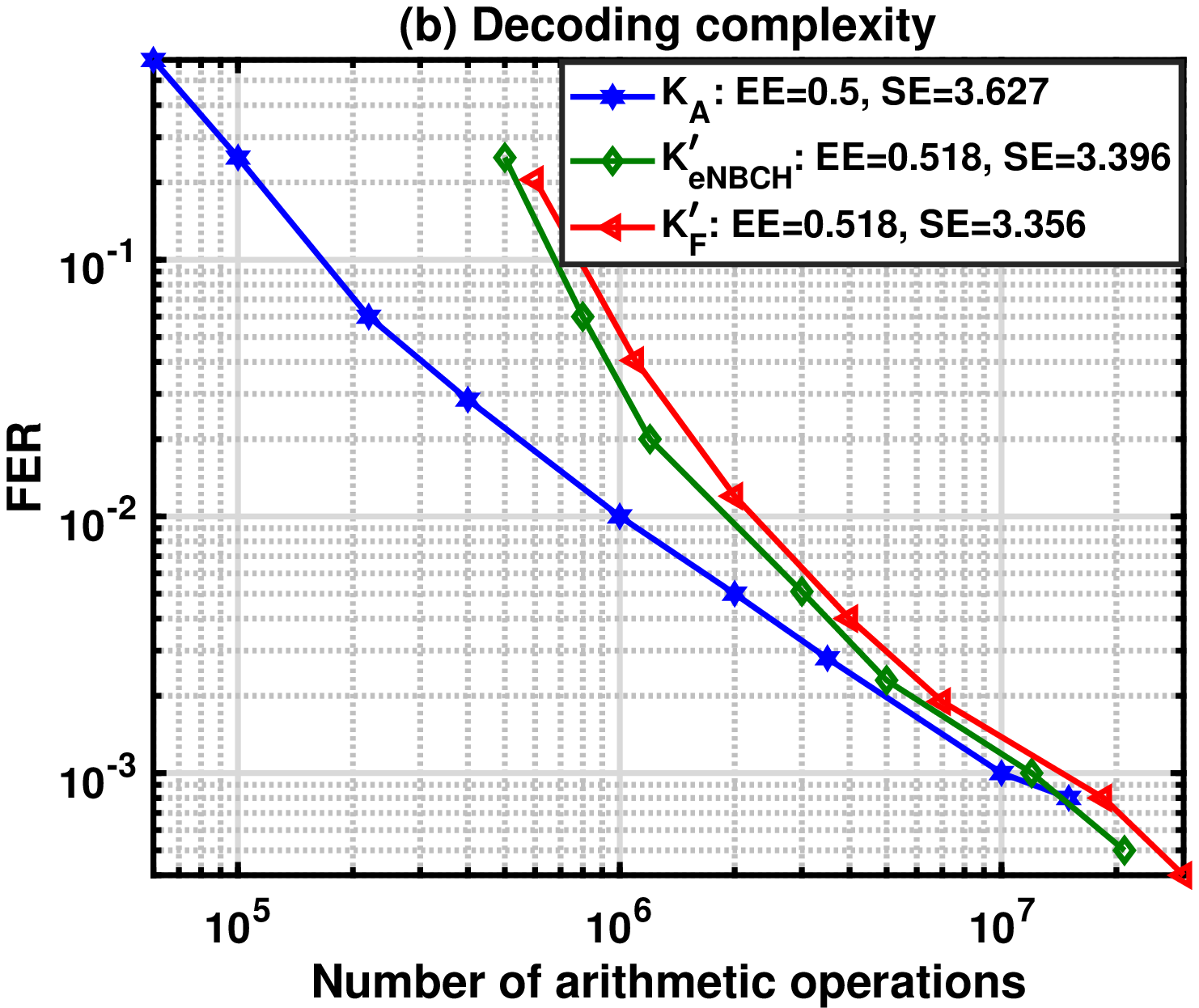}
        
     \end{subfigure}
        \caption{SCL decoding of polar subcodes with different kernels.}
        \label{fig:PerformanceComplexity}
\end{figure}

Fig. \ref{fig:Performance} illustrates the performance of the $(4096, 2048)$ polar (sub)codes constructed with  kernels $K^{\prime}_{\text{eNBCH}}$ and $K^{\prime}_{F}$ of size $16$ under SC and SCL decoders over the AWGN channel with BPSK modulation. \textcolor{black}{The construction is based on Monte-Carlo simulations.} Note that the column permutation \textcolor{black}{does not} alter the polarization behaviours of the kernel, so the performance of kernel $K$ is the same as the performance of kernel $K^{\prime}$. It can be seen that polar codes based on kernels $K^{\prime}_{F}$, $K^{\prime}_{\text{eNBCH}}$ provide significant performance gain compared to polar codes with Ar\i kan's kernel. \textcolor{black}{Indeed, kernel $K^{\prime}_F$ provides better performance compared to $K^{\prime}_{\text{eNBCH}}$, due to lower scaling exponent.} It can be observed that randomized polar subcodes \cite{Tsubcode} with $K^{\prime}_{eNBCH}$ under SCL with list size $L=8$ provides approximately the same performance as polar subcodes with Ar\i kan's kernel under SCL with $L=32$. 

Fig. \ref{fig:PerformanceComplexity} (a) shows the performance of the $(4096, 2048)$ polar subcodes constructed with $K^{\prime}_{\text{eNBCH}}$ and $K^{\prime}_F$ of size $16$ under SCL with different list sizes $L$ at \textcolor{black}{$Eb/No=1.25$ dB}. It can be observed that these kernels need a lower list size to achieve the same performance as $K_A$. Moreover, $K^{\prime}_F$ performs slightly better than $K^{\prime}_{\text{eNBCH}}$ with the same list size $L$.
Fig. \ref{fig:PerformanceComplexity} (b) shows the actual decoding complexity of the polar subcodes constructed with kernels $K^{\prime}_{\text{eNBCH}}$ and $K^{\prime}_F$ of size $16$ in terms of the number of arithmetical operations (summation and comparison) for different list sizes. It can be observed that $K^{\prime}_{\text{eNBCH}}$ provides better performance with the same decoding complexity for $\text{FER} \leq 8 \times 10^{-4}$ compared to the Ar\i kan's kernel.

\section{Conclusion}
In this paper, \textcolor{black}{we modified 
some good polarization kernels avaliable in the literature to reduce the computational complexity of their window decoder.} This modification, based on the  column permutations of the kernel, was applied to some  kernels constructed in the literature, e.g. eNBCH, 
and the results showed that the complexity of the window decoder for these modified  kernels was substantially lower as compared to the original ones.

\section*{ Acknowledgment}
The first author would like to thank Peter Trifonov, Grigorii Trofimiuk, Mohammad Rowshan,  Hanwen Yao and Lilian Khaw for very helpful discussion. The authors would like to thank the Editor, Khoa Le, and the anonymous reviewers for their constructive
comments.


%


\ifCLASSOPTIONcaptionsoff
  \newpage
\fi

\end{document}